\documentclass[prd,showpacs,showkeys,twocolumn]{revtex4}
\usepackage{bm}
\input{epsf}

\def\x{\left(\frac{r}{R}\right)}
\def\bet{\left(\frac{v_c}{c}\right)}
\begin{document}
\title{ 
Modelling galaxy halos using dark matter with pressure}
\author{Somnath Bharadwaj and Sayan Kar}
\email{sayan@cts.iitkgp.ernet.in, somnath@cts.iitkgp.ernet.in}
\affiliation{Department of Physics and Centre for Theoretical Studies 
\\ 
Indian Institute of Technology, Kharagpur 721 302, WB, India}
\begin{abstract}
We investigate whether a dark matter with substantial amounts of
 pressure, comparable in magnitude to the energy density, could be a
 viable candidate for the constituent of dark matter halos.  
We find that galaxy halos models, consistent with observations of 
 flat rotation curves, are  possible for a variety of equations of
 state  with anisotropic pressures. It turns out that the
 gravitational bending   of light rays  passing through such halos is
 very sensitive to the  pressure. We propose that combined
 observations of rotation curves and  gravitational lensing  can be
 used to  determine the equation of  state of the dark
 matter. Alternatively,   if the equation of state is known from other
 observations,  rotation  curves and gravitational  lensing  can
 together be used to test  General Relativity on galaxy scales.   
   
\end{abstract}


\maketitle

\section{Introduction}
The rotation curves  of spiral galaxies \cite{rc} are one of the most
direct probes of dark matter {\cite{dark}}. 
In these galaxies  neutral hydrogen (HI) clouds are observed at large  
distances from the center, much beyond the extent  of the luminous
matter.   The frequency shifts in the 21 cm HI
emission from these clouds allow their velocities  to be measured.
These clouds move  in circular orbits with velocity $v_c(r)$. 
The orbits are maintained by a balance of the centrifugal acceleration
$v^2_c/r$ and the gravitational  pull $G M(r)/r^2$ of the total mass
$M(r)$ contained within the orbit . This allows us to use the observed
rotation curves ($v_c$ as a function of $r$) to determine the mass
profile $M(r)=r v^2_c(r)/G$ of the galaxy. 

The rotational velocities  are  observed to increase near the center of the
galaxy and then remain nearly constant  at a value  $v_c \sim 200 \,
{\rm   km/s}$.  This leads to a mass  
profile $M(r)=r v_c^2/G$  where the mass within a distance $r$ from
the center of the galaxy increases linearly with $r$ even at large
distances where there is very little luminuos matter seen. This is
usually explained by postulating the existence of some  dark (invisible)
matter  distributed in a spherical  halo around the
galaxy. A modified  law of gravity  has  been proposed  as an 
alternative explanation {\cite{mond}}, we do not consider this here.  

There is at present no clear picture as to what constitutes the dark
matter in galaxy halos. The prevalent belief is that it  is a
pressureless medium (Cold Dark Matter) possibly 
made of weakly interacting massive particles  {\cite{cdm}}
(WIMPS). Here we consider the possibility that the dark matter in
galactic halos  has substantial pressure, comparable in magnitude to
the energy density.   

The Cold Dark Matter (CDM) model has been extremely successful in
explaining 
the observed CMBR anisotropies and large scale  structures in the 
universe {\cite{cdmcmb}}. The evolution of cosmological perturbations is
drastically modified in  models where the dark matter is an ideal
fluid with   significant amounts of pressure, and such models are
ruled out by these observations. Our study considers  the
possibility that the dark matter could be a medium which is not an
ideal fluid and has significant pressure. Such a situation is generic
if the dark matter is modelled as a scalar field (for references on
the use of scalar fields in modelling dark matter/dark energy  
see the papers cited in
{\cite{scalar}}), a
scenario which has come up quite often in the various attempts to
solve the problems being faced by the CDM model on small scales
\cite{cdmcusps} . This, and the fact that there has, till date, been
no direct detection of the CDM particles motivates us to explore the
possibility that the dark matter in galaxy halos may have pressure.  
In this work we do not restrict ourselves to any
particular model for the dark matter. We  consider  a general
situation where the pressure inside the halo is anisotropic and is
related to the energy density through a linear equation of state. 
Other attempts at interpreting  the flat rotation  
curves in the recent past are available in {\cite{lake}}. 

We  investigate whether it is possible to explain the
flat rotation curves seen in the outer parts of spiral galaxies using
dark matter which has pressure. The Newtonian analysis 
of the rotation curves outlined in the beginning of this section fails
when the pressure is comparable to the energy density,  and a 
relativisitic treatment is required. In the Newtonian analysis the
gravitational field inside the halo is described by a single
function, the gravitational potential, which  is completely
determined by the observations of the rotation curve. In the
relativisitic treatment the gravitational field inside  a 
static, spherically symmetric dark matter halo is described by the two 
unknown metric coefficients $g_{00}(r)$ and $g_{rr}(r)$. 
In section 2 of this paper we investigate to what extent the metric
inside the halo can be determined by observations of the rotation
curve.  We show that while $g_{00}(r)$ is completely determined by
just the rotation curve, inputs regarding the theory of gravity and
the dark matter equation of state are required to determine
$g_{rr}(r)$.    

In section 3. we calculate the gravitational lensing properties of the
dark matter halo. The usual analysis of gravitational lensing
{\cite{halolens}} does not take into account pressure. Here we analyse  photon
geodesics  inside the  halo and use these to calculate the
gravitational lensing deflection angles.  Subsequently,  we 
study to what extent the lensing properties of  dark matter halos
change when pressure is introduced. We investigate whether it is
possible to use combined observations of  rotation curves and
gravitational lensing to constrain the equation of state of the dark
matter in galaxy halos. In section 4. we discuss  our results and
present conclusions. 

\section{The geometry within the halo}
It is assumed that spiral galaxies are embedded in  static, spherically
symmetric halos of radius $R$. The gravitational field outside the
halo is described by the exterior  Schwarzschild metric with  
$M=R v^2_c/G$. The gravitational field inside the halo is represented
by two unknown functions $g_{00}(r)=e^{2 \Phi(r)}$ and $g_{rr}(r)=e^{2
  \lambda(r)} $ which parametrise the  metric, and the proper time
interval $d \tau$ is     
\begin{equation}
c^2 d \tau^2=e^{2 \Phi(r)} c^2 d t^2 - e^{2 \lambda(r)} d r^2 - r^2 \left[ d
  \theta^2 + \sin^2 \theta d \phi^2 \right] \,.
\end{equation}
We first investigate to what extent the gravitational field inside the
halo can be determined from  observations of the rotation
curves.

The HI clouds from which the frequency shifted 21
cm emission is observed are treated as test particles moving in
circular orbits, in the plane of the disk of the galaxy  under the
gravitational influence of the halo. It is assumed that the
HI clouds do not contribute significantly to the halo matter.   
We choose coordinates so that the disk of the galaxy is aligned with
the  equatorial plane of the halo ($\theta=\pi/2$). The orbits of the 
HI clouds have  two conserved quantities,  $E=e^{2 \Phi} c d t/d \tau$
and $L=r^2 d \phi/d \tau$ which, for circular orbits,  are related as 
$E^2=e^{2 \Phi} \left[ c^2 + L^2/r^2 \right]$. The HI clouds at
different radii will have different values of $E$ and $L$, and we
denote this by $E(r)$ and $L(r)$. 

A distant observer  in the plane of the disk receives HI
emission from a cloud located along a  line of sight  which  
is tangential to the orbit of the cloud. The observed shift in the
frequency of 
the radiation, when the cloud is moving away, is   $\Delta \nu
\approx-\nu_e L(r)/(c\,r)$, provided $e^{2  \Phi(r)} \sim 1$ and
$L(r)/(c\,r) \ll 1$.  Later in this paper we show that these two 
approximations  are  valid for our analysis.  An extra  angular  dependence   
comes into the frequency shift if the line of sight is at an angle
to the direction of motion of the HI cloud.  Observations of the
frequency shifts in the HI radiation show that, barring the
central parts of galaxies, the ratio $L(r)/(c \,r)$ is nearly  constant
at a value $L(r)/(c \, r)=v_c/c \sim 7\times 10^{-4}$. The ratio
$v_c/c$ being  extremely small, in our analysis, we retain only the
leading order terms in $v_c/c$, and ignore all the higher terms.

We next shift our attention to the equation of motion of the HI
clouds.  For circular orbits the geodesic equation reduces to 
\begin{equation}
\Phi^{'}(r) =\frac{ L^2(r)}{r^3 E^2(r)} \,.
\label{eq:a2}
\end{equation}
where $\Phi^{'}=d \Phi/dr$.
This can be  integrated to obtain
\begin{equation}
\Phi(r)=\int \left(\frac{L(r)}{r c}\right)^2 \frac{d r}{r} + constant \,.
\label{eq:a3}
\end{equation}
This equation allows us to determine the metric coefficient
$g_{00}(r)$ inside the halo using just the observed rotation 
curves. For a flat rotation  curve this gives us  
\begin{equation}
\Phi(r)=\left( \frac{v_c}{c} \right)^2 \left[ \ln(r/R)-1 \right]
\label{eq:a4}
\end{equation}
which gives the metric coefficient $g_{00}(r)$ to be   
\begin{equation}
g_{00}(r)=e^{2 \Phi(r)}=e^{- \frac{2 v^2}{c^2}}
\left(\frac{r}{R}\right)^{\frac{2 v_c^2}{c^2}} \,.
\label{eq:aa5}
\end{equation}
Here the constant of integration in equation (\ref{eq:a3}) has been
chosen so that the metric coefficient $g_{00}(r)$ matches the exterior
Schwarzschild metric at $r=R$. 

The rotation curve determines only one, namely
$\Phi(r)$ of the two unknown functions $\Phi(r)$ and $\lambda(r)$,
which are required to describe the gravitational field inside the
halo.  We also see that the metric coefficient $g_{00}(r)$ varies
very slowly with  $r$, and  $e^{2 \Phi(r)} \sim 1$, the  change being
less than $0.5 \%$ when $r$ changes from $0.01  R$ to $R$. The fact
that $g_{00}$ can indeed be obtained from the rotation curves, has been
{\em partially} noted in the papers in {\cite{lake}} and in some of the
references on halo dark matter {\cite{scalar}}.  

Next we address the question -- how to determine $\lambda(r)$? We proceed by
solving  Einstein equations for gravity $G_{\mu \nu}=(8 \pi G/c^4)
T_{\mu \nu}$ to determine $\lambda(r)$ inside the halo.   
Solving this requires the equation of state (or equivalently the
energy  momentum tensor $T_{\mu \nu}$) for the matter which makes up
the halo. We assume that the dark matter which makes up the halo is a
fluid with energy density $c^2 \rho(r)$, radial 
pressure $P_r(r)$, and tangential pressure ${P_T}(r)$ in the $\theta$
and $\phi$ directions.  Also, the gravitational field inside the halo
is weak $ie. \Phi(r) \sim \lambda(r) \sim (v_c/c)^2$ which allows us
to linearise the Einstein equations retaining only terms linear in
$(v_c/c)^2$.  Inside the halo, using equation (\ref{eq:a3}) for
$\Phi(r)$, the Einstein equations  reduce to three equations for
$\lambda(r)$ 
\begin{eqnarray}
\frac{(r \lambda)^{'}}{r^2}&=& \frac{4 \pi
G }{c^2} \rho  \label{eq:a5} \\
\frac{ (v_c/c)^2-\lambda }{r^2}
 &=&\frac{4 \pi G }{c^4}  P_r  \,  \label{eq:a6} \\ 
- \, \frac{\lambda^{'}}{r} 
&=&\frac{8 \pi G }{c^2} P_T \label{eq:a7}
\end{eqnarray}
which have to be solved  with the boundary condition
$\lambda(R)=(v_c/c)^2$ and the requirement $\rho(r) \ge 0$. 

In the absence of any pressure $(P_r=P_T=0)$ we recover the usual
Newtonian solution 
\begin{equation}
\lambda_N(r)=\bet^2 \hspace{1cm} {\rm and} \hspace{1cm}
\rho_N(r)=\frac{v^2}{4 \pi G r^2}
\label{eq:a8}
\end{equation}
This solution corresponds to  the singular isothermal sphere
which produces a flat rotation curve, 
We have solved equations (\ref{eq:a5}), (\ref{eq:a6}) and
(\ref{eq:a7})  under various assumptions on the equation of state. It
is convenient to express the results in terms of $\lambda_N$ and
$\rho_N(r)$ which are the soultions in the  absence of pressure.  

The most obvious possibility is to consider a medium with isotropic
pressure ($ie. \, P_r=P_T$). We find that the only solution where the
metric matches the Schwarzschild metric at $r=R$ and the  energy
density is positive has $P_r=P_T=0$.  

We next consider two possible anisotropic equations of state
A. $P_r=w_r \rho c^2$ where we use eq. (\ref{eq:a7}) to determine
$P_T$, and   
B. $P_T=w_T \rho c^2$ where we use eq. (\ref{eq:a6}) to determine
$P_r$. As mentioned in the Introduction, such anisotropic pressures
can indeed arise in the energy--momentum tensor of a real
or complex scalar field. We recall that the energy momentum tensor 
for a real scalar field in a potential $V (\phi)$ is given as :

\begin{equation}
T_{ij}^{\phi} = \partial_i \phi \partial_j \phi -\frac{1}{2}g_{ij} \left [
\partial^k \phi \partial_k \phi + 2V(\phi) \right ] 
\end{equation}

When the gradient terms are small, and the scalar field is frozen at
a potential minimum $\phi_{min}\neq 0$, it generates   
isotropic, negative pressures  $P=-\rho  c^2=-V(\phi_{min})$.   
On the other hand, the pressure is anisotropic {\it i.e. the radial
pressure $T_{rr}$ is different from the tangential  pressure
$T_{\theta \theta}$}, if the radial derivative of the scalar field
makes a significant contribution to the energy momentum tensor. 
Various scalar field configurations, including some with anisotropic
pressures (for example, Matos and Nunez, astro-ph/0303455 in {\cite{scalar}}), 
have been used extensively in cosmology to model the
dark matter and the dark energy. 
  
We now discuss the results for the two cases (A) and (B) mentioned above. 

\subsection{$P_r=w_r \rho c^2$}
We find that solutions exist for $w_r \ge -1$. If we match 
match $\lambda(r)$ to the Schwarzschild metric at the outer edge of
the halo, the density turns out to be negative for $w_r<-1$ and this
range is ruled out. The allowed solutions are tabulated below.  

\vspace{.3in}
\begin{widetext}
\large{
\begin{center}
\begin{tabular}{|c|c|c|c|}
\hline
$w_r$ &  $\lambda(r)=\lambda_N \times$ & $\rho(r)=\rho_N(r) \times$ &
$P_T(r)=(c^2/2) \rho_N(r) \times$ \\
\hline
$>-1$ &  $\frac{1}{1+w_r}\left[ 1 + w_r  \x^{\frac{-w_r}{1+w_r}} \right] $ &
$\left( \frac{1}{1+w_r}\right)\left[ 1 + 
  \frac{w_r}{1+w_r}  \x^{\frac{-w_r}{1+w_r}}  \right] $& 
$\left(\frac{w_r}{1+w_r} \right)^2   \x^{\frac{-w_r}{1+w_r}} $\\
\hline
$-1$ &$\left[1 -  \ln(r/R) \right]  $ & $ \ln \left (\frac{R}{r}\right ) $& $1$ \\
\hline 
\end{tabular}
\end{center}
}
\end{widetext}

\subsection{$P_T=w_T \rho c^2$}
We find that solutions exist for $w_T > -1/2$. For $w_T=-1/2$ we
have an absurd situation where $\lambda(r)=0$, and the density is
negative if $w_T<-1/2$. The  solution for the allowed range of $w_T$is
shown below  
\vspace{.3in}

\begin{widetext}
\large{
\begin{center}
\begin{tabular}{|c|c|c|c|}
\hline
$w_T$ &  $\lambda(r)=\lambda_N \times$ & $\rho(r)=\rho_N(r) \times$ &
$P_r(r)=c^2 \rho_N(r) \times$ \\
\hline
$>-1/2$ &  $\x^{\frac{-2 w_T}{1+2 w_T}} $ & $\frac{1}{1+2w_T}
\x^{\frac{-2 w_T}{1+2 w_T}} $ & $\left[1-\x^{\frac{-2 w_T}{1+2 w_T}}
  \right]$ \\ \hline
\end{tabular}

\end{center}
}
\end{widetext}

\section{\bf Gravitational lensing}
We now move on  to calculate the bending  of  light  by the
 halos.  A light ray  which goes past the halo without entering it
 propagates entirely in a Schwarzschild metric. The light ray is
 deflected  by an angle  $\Delta=4GM/c^2 b=4 (v_c/c)^2 (R/b)$ where
 $b$  is the impact parameter. and in this  case $b  \ge R$. The
 bending of a light ray which passes through the 
 halo is determined by  the metric inside the halo and this depends on
 the   equation of state of the dark matter. 

 The light ray is assumed to lie in the 
$\theta=\pi/2$ plane. Null-geodesics  in the halo metric 
are completely determined by the two conserved quantities  $E$ and $L$
defined in section 2. The trajectory of the light ray is determined by
the following equation {\cite{Shutz}}  for $u(\phi) = \frac{1}{r(\phi)}$ 

\begin{equation}
\left ( \frac{du}{d\phi} \right )^2 = e^{-2\left (\Phi + \lambda\right )
} \left ( \frac{E}{L} \right )^2 - e^{-2\lambda} u^2  
\label{c1}
\end{equation}
Retaining only terms to order $\sim (v/c)^2$, and using
$1/b^2=E^2/L^2$  we have 
\begin{equation}
\left ( \frac{du}{d\phi} \right )^2 = \left ( \frac{1}{b^2} - u^2
\right ) - 2 \left [ \left (\Phi + \lambda \right ) \frac{1}{b^2} -
\lambda u^2 \right ]
\label{eq:c2}
\end{equation}
It is convenient to replace $u$ by a new variable $y$ where
$y=u+\alpha(y)$ and 
\begin{equation}
\alpha (y) = \frac{\left (\Phi + \lambda \right )/b^2 - \lambda
  y^2}{y}
\label{eq:c3}
\end{equation}
We can now integrate (\ref{eq:c2}) to obtain 
\begin{equation}
\phi = \int_0^{1/b} \left (1-\frac{d\alpha}{dy} \right ) \left [
  \frac{1}{b^2} - y^2 \right ]^{-\frac{1}{2}} dy  
\label{eq:c4}
\end{equation}
In equation (\ref{eq:c4}) the limit $y=0$ is when the 
 in-coming photon is very far away from the center of the
halo,  $y=1/b$ is when the photon is closest to the center of the halo
and $\phi$ is the change in the angle subtended by the photon between
these two positions.  Setting $\alpha(y)=0$ allows us to calculate the
 photon trajectory in the absence of any gravitation field, whereas
 the term involving $\alpha$ is the change produced due to the
 gravitational field of the halo. This gives the  deflection
 angle to be 
\begin{equation}
\Delta = -2 \int_0^{1/b} \frac{d\alpha}{dy} \left (\frac{1}{b^2} - y^2
\right )^{-\frac{1}{2}} dy 
\label{eq:c5}
\end{equation}
where the factor of $2$ arises from the contributions of  the
in-coming and the outgoing photon trajectory. 

The integral in eq. (\ref{eq:c5}) is evaluated in two parts where we
use the Schwarzschild metric to calculate $\alpha$ outside the halo (
$1/y>R$) and we use the solutions obtained in Section 2. inside the
halo $(1/y<R)$. The results for the deflection angle
$\Delta$ are presented in units of $(v_c/c)^2$ using a variable
$\delta=\Delta/(v_c/c)^2$. We recover the familiar result $\delta=4(R/b)$
corresponding to the exterior Schwarzschild geometry when the impact
parameter is larger than halo radius ( $b\ge R$).   In the situation
when the light ray does enter the halo $(b < R)$, eq. (\ref{eq:c5})
can be evaluated analytically using Mathematica for both the equations
of state considered in Section 2. The resulting expressions are rather 
lengthy and we do not present them here. In the absence of pressure
we recover the familiar result $\delta=2 \pi$ for an infinite,
isothermal halo.  We recover this result from eq. 
(\ref{eq:c5}) in the limit $(b/R) \rightarrow 0$.  This can
be interpreted as either the halo raduis becoming very large for a
fixed  impact parameter,  or the impact parameter
approaching very close to the center of the halo with the halo radius
remaining constant.  The results for the deflection angle in the
presence of pressure, under the two different equations of state
assumed in section 2. are shown in figures 1 and 2. 

\begin{figure}

\centerline{\epsfxsize=3.25in\epsffile{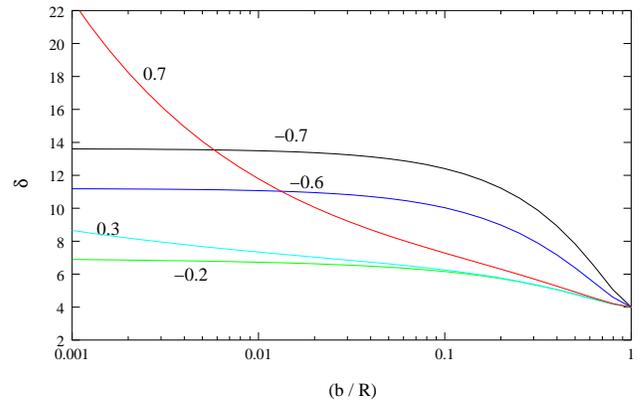}}

\caption{This shows the
  deflection angle $\Delta=(v_c/c)^2 \delta$ as a function of the
  impact parameter $b$ for a light ray passing through the dark matter
  halo of a galaxy which has  a flat rotation curve with speed $v_c$. 
The equation of state of the dark matter is assumed to be $P_r=w_r
  \rho c^2$ and the values of $w_r$ for which the deflection angles
  have been calculated are shown in the figure.}  

\end{figure}

\vspace{0.5in}

\begin{figure}

\centerline{\epsfxsize=3.25in\epsffile{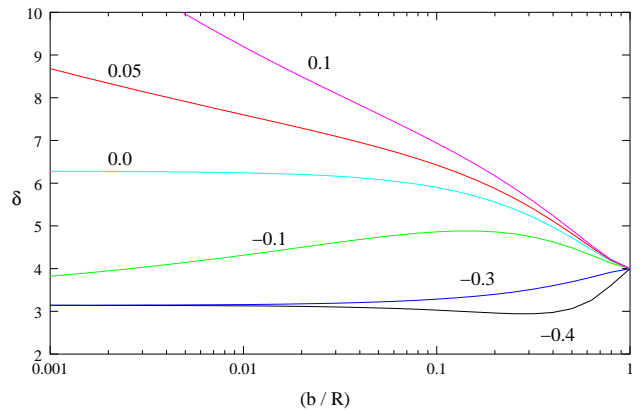}}

\caption{This figure is the same as
 Figure 1 except that equation of state of the dark matter is assumed
  to be $P_T=w_T   \rho c^2$}

\end{figure}

\vspace{.1in}

\section{Results and Conclusions}
We have addressed the question if the dark matter in the halos of
spiral galaxies could have significant amounts of pressure, which in
general could be anisotropic.  This requires a relativistic treatment 
where the gravitational 
field inside the halo is described by the two metric coefficients
$g_{00}(r)$ and 
$g_{rr}(r)$. We find that observations of the rotation curve fully
determine $g_{00}(r)$ irrespective of the equation of state of the
dark matter. As a matter of fact $g_{00}(r)$ is determined without any
reference to the theory for gravity, the only assumption being that
the gravitational effects can be represented through the space-time
metric.      

Determining $g_{rr}(r)$ requires us to assume a theory for gravity,
Einsteins's theory of General Relativity in this case, and an equation
of state for the radial pressure $P_r$ and the tangential pressure
$P_T$ of the dark matter. We consider two possibilities (A>)
$P_r=w_r \rho c^2$ with $P_T$ being determined from Einstein's
equations and (B.) $P_T=w_T \rho c^2$ with $P_r$ being determined from
Einstein's equations. It may be noted that there is no solution
matched to the Schwarzschild exterior metric at the boundary of the
halo $(r=R)$ if the pressure is assumed to be isotropic $(P_r=P_T)$.  

Solutions matched to a Schwarzschild exterior solution exist for 
(A.) $w_r >-1$ and (B.) $w_T>-1/2$. Making use of the calculated 
$g_{rr}(r)$ we determine the density  and pressure profiles for the
halos. The density profile is shallower than the $r^{-2}$ Newtonian
density profile only for $-1/2<w_T<0$.   

For $P_r=-\rho c^2$ we have an interesting situation where
$g_{rr}(r)=-g_{00}^{-1}(r)$ both inside and outside the halo. The
density also goes to zero at the boundary,  unlike in all the other
cases.  
It should be noted that we do not attempt to match the density and
pressures at the boundary of the halo, and in fact it is not possible
to do this for a flat rotation curve and a Schwarzschild exterior. 

We next investigate if it is possible to constrain $g_{rr}$ or
equivalently the equation of state using gravitational lensing. The 
bending of light rays was calculated for the cases (A.) and (B.) and
the results are summarised below. 

(A.) For $\mid w_r\mid >0$ the bending angle is always larger than
the case where there is no pressure. When $w_r>0$, the deflection
angle 
progressively increases as the halo radius $R$ is increased keeping
the impact parameter $b$ fixed, and it diverges in the limit   $(b/R)
\rightarrow 0$.  On the other hand, for $w_r<0$ the behaviour of the
deflection angle is similar to  the situation when there is no
pressure, except that in the limit $(b/R) \rightarrow 0$ it saturates at
a value larger than  $\Delta_N=2 \pi (v_c/c)^2$ which is  the
deflection for an infinite halo with no  pressure. 

(B.) The generic features of the deflection angle  are different for
the two cases $w_T>0$ and $w_T<0$. For $w_T>0$,  the deflection angle
is larger than $\Delta_N$,  and it increases progressively as
the halo radius is increased, diverging in the limit
$(b/R)\rightarrow 0$.  For  $w_T<0$,  the deflection angle is less than
$\Delta_N$ and it converges to around $0.5 \Delta_N$  in the limit
$(b/R) \rightarrow 0$.  

We find that  in case (B. ) with $w_T<0$,  the only situation where 
the deflection angle is less than $\Delta_N$,  the  density profile is
shallower than  $r^{-2}$. The density  has a  $r^{-2}$ behaviour or is
steeper in all the other the cases. It may be worthwhile to remind the 
reader here that in all the cases considered here  both the 
energy density and the pressure contribute to the bending of light.  
We next ask if there is any pattern seen in the form  of
$g_{rr}(r)$ (or effectively $\lambda(r)$) and the behaviour of the
deflection angle. We find that $w_T<0$ is the only case where 
$\lambda(r)$   is a  monotonically increasing function of $r$. This
leads to a picture where the gravitational deflection angle is
$\Delta_N$  in the absence of pressure where $\lambda(r)$
is a constant, the deflection is more than $\Delta_N$ if $\lambda(r)$
is a monotonically decreasing function of $r$ and the deflection is
less than $\Delta_N$ if  $\lambda(r)$
is a monotonically increasing function of $r$. 

In conclusion, the flat rotation curves observed in the outer parts of
spiral galaxies can be equally well explained by a halo which is made
up of dark matter with anisotropic pressures. The bending of light
rays passing through the halo is found to be highly sensitive to the
pressure content of the dark matter. This holds the possibility that
combined observations of rotation curves and gravitational lensing can
be used to determine the equation of state of the dark matter in
galactic halos. Alternatively, if the equation of state of the dark
matter is determined from other observations, then combined
observations of rotation curves and gravitational lensing can be used
to test Einstein's theory of General Relativity on galaxy scales.

\newpage

\end{document}